\documentclass[prl,twocolumn,floatfix,bibnotes]{revtex4-1}

\usepackage{physics}
\usepackage{gensymb}
\usepackage{graphicx}
\usepackage{amsmath}
\usepackage[hidelinks]{hyperref}
\usepackage{newfloat}
\DeclareFloatingEnvironment[name={FIG. S}]{suppfigure}

\begin{document}

\title{Bright soliton to quantum droplet transition\\ in a mixture of Bose-Einstein condensates}
\author{P. Cheiney, C. R. Cabrera, J. Sanz, B. Naylor, L. Tanzi, and L. Tarruell}\email{Electronic address: leticia.tarruell@icfo.eu}
\affiliation
{ICFO -- Institut de Ci\`{e}ncies Fot\`{o}niques, The Barcelona Institute of Science and Technology, 08860 Castelldefels (Barcelona), Spain}

\begin{abstract}
Attractive Bose-Einstein condensates can host two types of macroscopic self-bound states of different nature: bright solitons and quantum liquid droplets. Here, we investigate the connection between them with a Bose-Bose mixture confined in an optical waveguide. We develop a simple theoretical model to show that, depending on atom number and interaction strength, solitons and droplets can be smoothly connected or remain distinct states coexisting only in a bi-stable region. We experimentally measure their spin composition, extract their density for a broad range of parameters and map out the boundary of the region separating solitons from droplets.
\end{abstract}

\date{\today}
\maketitle

Self-bound states are ubiquitous in nature, and appear in contexts as diverse as solitary waves in channels, optical solitons in non-linear media and liquid He droplets \cite{SolitonBook,Malomed2005, Barranco2006}. Their binding results from a balance between attractive forces, which tend to make the system collapse, and repulsive ones, which stabilize it to a finite size.

Bose-Einstein condensates (BECs) with attractive mean-field interactions constitute ideal model systems to explore in the same setting self-bound states stabilized by repulsive forces of different classes. On the one hand, bright solitons in optical waveguides have been observed with $^{7}$Li \cite{Strecker2002, Khaykovich2002, Medley2014}, $^{85}$Rb \cite{Cornish2006, Marchant2013, McDonald2014} and $^{39}$K atoms \cite{Lepoutre2016}. These matter-wave analogues of optical solitons are stabilized against collapse by the dispersion along the unconfined direction, which is a (single-particle) kinetic effect. On the other hand, quantum droplets -- self-bound clusters of atoms with liquid-like properties -- have been recently demonstrated with $^{164}$Dy \cite{Kadau2016, Ferrier2016a, Ferrier2016b, Schmitt2016} and $^{166}$Er atoms \cite{Chomaz2016}, and in mixtures of $^{39}$K BECs \cite{Cabrera2017}. In this case, the repulsive force preventing the collapse stems from quantum fluctuations, and has a quantum many-body origin \cite{Petrov2015}.

Bright solitons and quantum droplets are \emph{a priori} distinct states which exist in very different regimes. Solitons require the gas to remain effectively one-dimensional, which limits their maximal atom number \cite{Perez1998, Salasnich2002, Carr2002}. In contrast, droplets are three-dimensional solutions that exist even in free space and require a minimum atom number to be stable \cite{Petrov2015,Bisset2016,Wachtler2016a,Schmitt2016, Chomaz2016, Cabrera2017}. Up to now, quantum droplet experiments focused exclusively on systems where solitons were absent, enabling an unambiguous identification of the droplet state. Therefore, they could not provide any insights on their connections to solitons.

\begin{figure}[!t]
\centering
\includegraphics[clip]{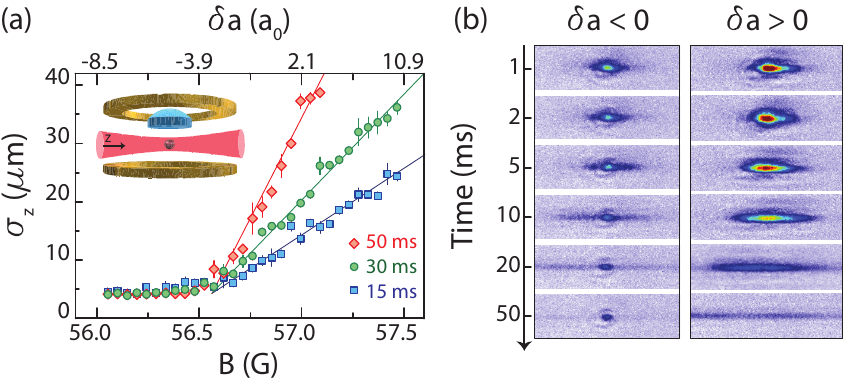}
\caption{Self-bound states. (a) Gaussian $1/\mathrm{e}$ width $\sigma_z$ of the mixture as a function of the magnetic field $B$ (corresponding to different values of $\delta a$), for various evolution times after release in the optical waveguide (inset). For $B<56.6$ G the system becomes self-bound and $\sigma_z$ saturates to the imaging resolution. Solid lines are linear fits to the data in the expanding regime and error bars denote the standard deviation of 10 independent measurements. (b) Typical \emph{in situ} images for increasing evolution times, corresponding to a self-bound state (expanding gas) in the attractive (repulsive) regime with $\delta a<0$ ($\delta a>0$).}\label{fig1}
\end{figure}

\begin{figure*}[!t]
\centering
\includegraphics[scale=1,clip]{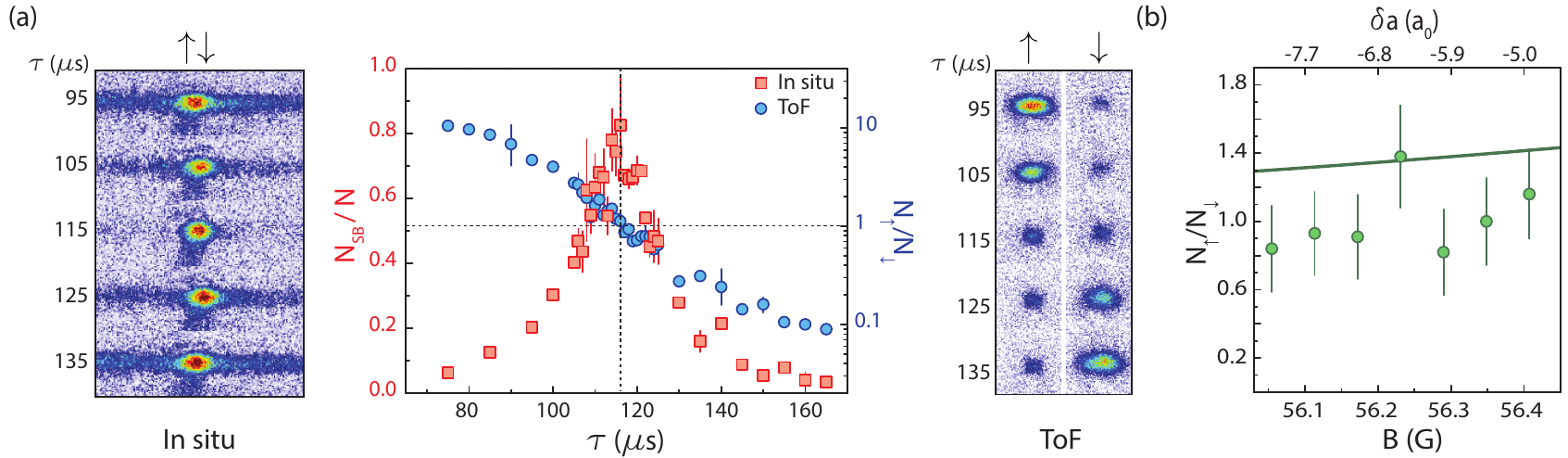}
\caption{Spin composition. (a) Left panel: \textit{in situ} images of the mixture for various rf pulse times $\tau$ and $B=56.35(1)$ G. Away from an optimal value the density profile is bi-modal, with a self-bound state surrounded by atoms of the excess component. Central panel: fraction of self-bound atoms $N_{\mathrm{SB}}/N$ (red squares) and spin composition $N_{\uparrow}/N_{\downarrow}$ (blue circles) as a function of $\tau$. Error bars denote the standard deviation of 4 measurements. Right panel: corresponding time-of-flight (ToF) Stern-Gerlach analysis of the spin composition. (b) Optimal ratio $N_{\uparrow}/N_{\downarrow}$ as a function of magnetic field $B$. Error bars correspond to the confidence interval of the fit \cite{NoteSupplementary}. The solid line depicts the theoretical prediction $N_{\uparrow}/ N_{\downarrow} =\sqrt{a_{\downarrow\downarrow}/ a_{\uparrow\uparrow}}$.}\label{fig2}
\end{figure*}

In this Letter, we bridge this gap by exploring a system that can host both bright solitons and quantum droplets: a mixture of two BECs in an optical waveguide. We observe that, as soon as the mean-field interactions become effectively attractive, self-bound states of well-defined spin composition appear. We show theoretically that their nature evolves from soliton-like to droplet-like upon increase of the atom number. Depending on the interaction strength, both regimes can be smoothly connected, or remain distinct states that coexist only in a bi-stable region. We determine experimentally their density for a broad range of atom numbers and interaction strengths, and map out the boundary of the bi-stable region that separates bright solitons from quantum droplets.

We perform experiments with a mixture of  $^{39}$K BECs in Zeeman states $\ket{\uparrow}\equiv\ket{m_F=-1}$ and $\ket{\downarrow}\equiv\ket{m_F=0}$ of the $F=1$ hyperfine manifold. The optical waveguide is created by a red-detuned optical dipole trap of radial trapping frequency $\omega/2\pi=109(1)$~Hz, see inset of Fig. 1(a). The system is imaged \emph{in situ} with a spatial resolution of the order of the harmonic oscillator length $a_{\mathrm{ho}}=\sqrt{\hbar/m\omega}\simeq1.5\,\mu$m, with $\hbar$ the reduced Planck constant and $m$ the mass of $^{39}$K. We exploit a phase-contrast polarization scheme \cite{Bradley1997} to image both states with the same sensitivity \cite{Cabrera2017}. The interactions are tuned  \emph{via} magnetic Feshbach resonances and parameterized by the intra- and inter-component scattering  lengths $a_{\uparrow\uparrow}, a_{\downarrow\downarrow}>0$ and $a_{\uparrow\downarrow}<0$ \cite{Roy2013}. The overall mean-field interaction is proportional to $\delta a= a_{\uparrow \downarrow }+\sqrt{ a_{\uparrow \uparrow} a_{\downarrow \downarrow }}$, which is attractive for \mbox{$B<56.84$ G} \cite{NoteSupplementary}.

The experiment starts with a pure BEC in state $\ket{\uparrow}$ confined in a crossed optical dipole trap. A radio-frequency (rf) pulse is used to prepare a controlled mixture of the two components at $B\sim57.2$ G, where $\delta a > 0$ and the system is in the miscible regime \cite{Hall1998}. Subsequently the magnetic field is ramped down at a constant rate of $11.75$ G/s while reducing the longitudinal confinement. The latter is removed in 5 ms at the final magnetic field, leaving the system unconfined along the $z$ direction.

Fig. 1(b) shows typical \emph{in situ} images of the time evolution of the mixture after release in the optical waveguide. Fig. 1(a) displays its longitudinal size $\sigma_z$ as a function of magnetic field, for three different evolution times. In the repulsive regime ($\delta a> 0$) $\sigma_z$ increases with $\delta a$, reflecting the increase of the released energy of the gas. In contrast, in the attractive regime ($\delta a< 0$) the absence of expansion indicates the existence of self-bound states. Experimentally, we only observe this behavior below $\delta a\sim-2 a_0$, where  $a_0$ denotes the Bohr radius. As in ref. \cite{Lepoutre2016}, we attribute this effect to the initial confinement energy of the system.

The observed self-bound states are intrinsically composite objects, involving both $\ket{\uparrow}$ and $\ket{\downarrow}$ atoms. To probe this aspect, we prepare mixtures of different compositions by varying the rf pulse time $\tau$. Large population imbalances between the two states result in bi-modal density profiles in the \emph{in situ} images, see left panel of Fig. 2(a). They consist of a self-bound state surrounded by a wider and expanding cloud of atoms of the excess component. We find that the fraction of self-bound atoms is maximized for an optimal pulse time, see central panel.

To determine its spin composition we perform a complementary set of measurements, modifying the detection sequence. We dissociate the self-bound state by increasing the magnetic field to the repulsive regime ($B\sim 57.3$ G) in $1$ ms, similary to ref. \cite{Kadau2016}. We then measure the atom number per spin component $N_{\uparrow}$ and $N_{\downarrow}$ \emph{via} Stern-Gerlach separation during time-of-flight expansion, see right panel. We extract the optimal composition as a function of $B$ by combining the \textit{in situ} and time-of-flight measurements, see Fig. 2(b). The interaction energy of the system is minimized by maximizing the spatial overlap of the two components \cite{Petrov2015, Petrov2016}. The theoretical prediction assuming that both occupy the same spatial mode yields $N_{\uparrow}/N_{\downarrow}=\sqrt{a_{\downarrow\downarrow}/a_{\uparrow\uparrow}}$ (solid line), which is in fair agreement with the data.

To clarify the nature of the self-bound states and their relation to the well-known bright soliton and quantum droplet limits, we perform a theoretical analysis of the system. It is based on the extended Gross-Pitaevskii equation (eGPE) proposed in ref. \cite{Petrov2015}, which includes the effect of quantum fluctuations through an additional repulsive term. Its derivation assumes explicitly that the two components occupy the same spatial mode $\Psi_{\uparrow}=\sqrt{n_{\uparrow}}\phi$ and $\Psi_{\downarrow}=\sqrt{n_{\downarrow}}\phi$, where $n_0=n_{\uparrow}+n_{\downarrow}$ is the total peak density of the system and $n_{\uparrow}/n_{\downarrow}=\sqrt{a_{\downarrow\downarrow}/ a_{\uparrow\uparrow}}$. The system is then described by
\begin{equation}
i\hbar\dot{\phi}=\left[\left(-\frac{\hbar^2}{2 m}\nabla^2+V_{\mathrm{trap}}\right)+\alpha\, n_0\left|\phi\right|^2+\gamma\, n_0^{3/2} \left|\phi\right|^3\right]\phi,\nonumber \label{eGPE}
\end{equation}
where $V_{\mathrm{trap}}$ denotes the waveguide confinement, and $\alpha \propto \delta a$ and  $\gamma \propto (\sqrt{a_{\uparrow\uparrow}a_{\downarrow\downarrow}})^{5/2}$ are functions of the magnetic field \cite{NoteSupplementary}. Note that although this equation bears strong similarities with the cubic-quintic non-linear Schr\"{o}dinger equation employed in optics to describe high-order material non-linearities \cite{Malomed2005}, the repulsive term has an unusual quartic dependence. This is the scaling corresponding to quantum fluctuations in three-dimensional condensates  \cite{Lee1957}, which is the regime explored experimentally \cite{DropletDimensionalityNote}.

\begin{figure}[t]
\centering
\includegraphics[scale=0.98,clip]{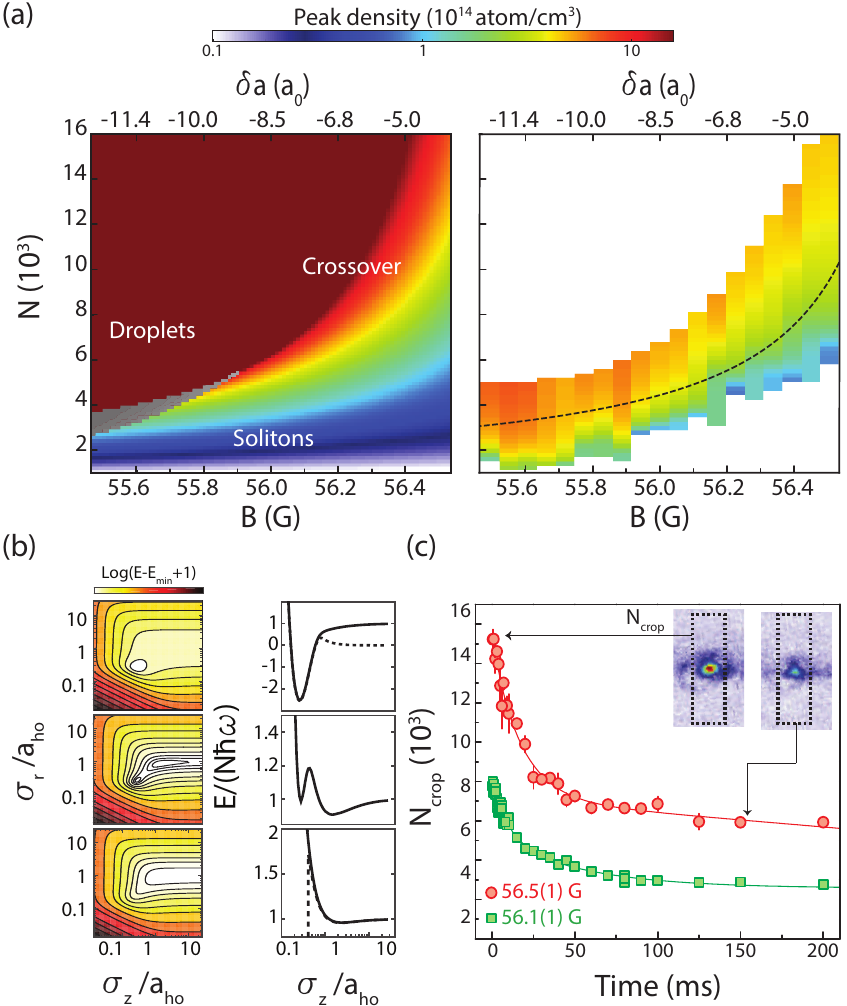}
\caption{Soliton-to-droplet density diagram. (a) Left panel: ground state peak density as a function of atom number $N$ and magnetic field $B$ computed numerically from the eGPE. Solitons and droplets are distinct solutions, which coexist in a bi-stable region (gray area) and become smoothly connected in the crossover above $B_c\sim 55.85$ G. Right panel: peak density extracted from the decay of the self-bound atom number, see (c). Self-bound states are stabilized by beyond mean-field effects well above the mean-field collapse threshold for composite bright solitons (dashed line). (b) Left panel: energy $E$ of the Gaussian ansatz as a function of the radial $\sigma_r$ and longitudinal $\sigma_z$ sizes, for $B = 55.6$ G and $N=6000$ (top, droplet), $N=3700$ (center, bi-stable region) and $N=2500$ (bottom, soliton). Right panel: one-dimensional cuts along $\sigma_z$, for $\sigma_r$ that minimizes $E$. All panels, solid lines: complete model; top panel, dashed line: no optical waveguide; bottom panel, dotted line: no quantum fluctuations. (c) Evolution of the self-bound atom number $N_{\mathrm{crop}}$, determined from the zeroth moment of the cropped region (insets), as a function of time $t$. Solid lines: empirical fit for extracting the decay rate \cite{NoteSupplementary}. Error bars: standard deviation of 4 measurements.}\label{fig3}
\end{figure}

We compute the ground state of the system by solving numerically the eGPE \cite{NoteSupplementary}. The left panel of Fig. 3(a) depicts its peak density $n_0$ as a function of the total atom number $N=N_{\uparrow}+ N_{\downarrow}$ and magnetic field $B$ (equivalently, interaction strength $\delta a$). For large attraction we find two distinct behaviors: a high-density solution ($n_0\sim 10^{16}$ atoms/cm$^3$) for large $N$, and a low-density one ($n_0\sim 10^{13}$ atoms/cm$^3$) for small $N$. In between, the gray region corresponds to a bi-stable regime where both solutions are possible. Its boundaries are signalled by a discontinuity of the density, as in a first order liquid-to-gas phase transition. This behavior disappears above a critical magnetic field ($B_c \sim 55.85$ G for our experimental confinement). Beyond $B_c$ the system supports a single solution whose density increases progressively with $N$.

For all parameters considered in Fig. 3(a), we find that the density profile of the system is well approximated by a Gaussian.  To gain further insight on the phase diagram, we thus perform a variational analysis of the eGPE \cite{Perez1997} introducing the ansatz $\phi=\mathrm{e}^{-r^2/2\sigma_r^2-z^2/2\sigma_z^2}$ \cite{NoteSupplementary}. Fig. 3(b) displays the energy landscapes obtained at a fixed magnetic field $B=55.6$ G $<B_c$. For small values of $N$ (bottom row), the energy has a single minimum corresponding to a dilute and elongated cloud: a composite bright soliton. Its radial size $\sigma_r$ corresponds to the harmonic oscillator length $a_{\mathrm{ho}}$ and its longitudinal size $\sigma_z$ and energy $E$ are similar to those obtained in a mean-field treatment without quantum fluctuations (bottom right panel, dotted line). For large values of $N$ (top row) the minimum corresponds to a dense and isotropic solution with $\sigma_r\ll a_{\mathrm{ho}}$: a quantum liquid droplet. Its properties are not affected by the trapping potential, and it exists in its absence (top right panel, dashed line). In the bi-stable region (central row) both composite bright solitons and liquid droplets exist simultaneously. Above the critical magnetic field $B_c$ a crossover takes place, with a single solution which evolves from soliton-like to droplet-like upon increasing the atom number. A related behavior, involving a bi-stable region and a crossover regime, has been studied in harmonically trapped dipolar gases \cite{Wachtler2016a,Bisset2016,Schmitt2017}. In this case, the low- and high-density solutions correspond to a BEC and a quantum droplet.

\begin{figure*}[!t]
\centering
\includegraphics[scale=0.9,clip]{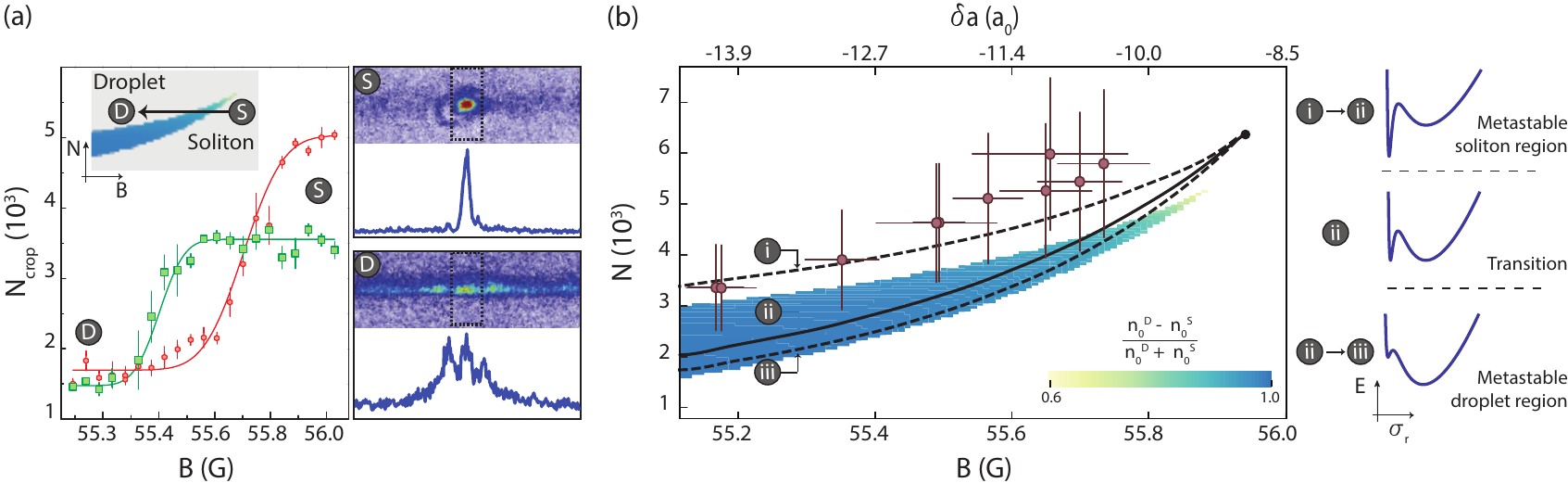}
\caption{Soliton-to-droplet transition. (a) Left panel: Atom number in the self-bound region $N_{\mathrm{crop}}$ as a function of magnetic field $B$ when approaching the bi-stable region from the soliton regime, see inset. Top right panel: initial soliton image (S) and corresponding doubly-integrated density profile. Bottom right panel: fragmentation observed when entering the droplet regime (D). (b) Measured fragmentation point vs. $N$ and $B$. Error bars: systematic error in $N$ and magnetic field width of the fragmentation curve \cite{NoteSupplementary}. Colored area: bi-stable region computed numerically from the eGPE. Lines: variational model, indicating the boundaries of the bi-stable region (dashed) and the transition line where solitons and droplets have identical energies $E$ (solid). Insets: sketch of $E$ vs. $\sigma_z$ for the metastable soliton and droplet regions and the transition line.}\label{fig4}
\end{figure*}

We explore experimentally the phase diagram of the system preparing self-bound states at different interaction strengths, starting from the high atom number regime. We observe that their atom number decreases in time due to inelastic processes, see Fig. 3(c). For our experimental parameters these are completely dominated by three-body recombination in the $\downarrow\downarrow\downarrow$ channel \cite{NoteSupplementary}. We model the decay of the self-bound atom number using the simplified rate equation $\dot{N}/N=-K^{\mathrm{eff}}_3 \langle n^2\rangle$, where $\langle n^2\rangle$ is the total mean square density and $K_3^{\mathrm{eff}}$ an effective three-body loss coefficient. The model assumes that the $\ket{\downarrow}$ losses are accompanied by the expulsion of $\ket{\uparrow}$ atoms from the self-bound state in order to maintain the value of $N_{\uparrow}/ N_{\downarrow}$ constant \cite{NoteSupplementary}.

Similarly to recent experiments on dipolar $^{166}$Er droplets \cite{Chomaz2016}, we extract the density of the self-bound state by measuring the decay of its atom number. The latter allows us to map out the density as a function of $N$ from a single decay curve, overcoming the limits set by the imaging resolution. The right panel of Fig. 3(a) displays the determined peak densities as a function of atom number and magnetic field.  Interestingly, a large fraction of the measurements lies well above the mean-field bright soliton collapse threshold. At the theoretical optimum for $N_{\uparrow}/N_{\downarrow}$ it corresponds to the condition $N_c= 0.6268 \,a_{\mathrm{ho}} \left(1+\sqrt{a_{\downarrow\downarrow}/a_{\uparrow\uparrow}}\right)^2/\left(2|\delta a| \sqrt{a_{\downarrow\downarrow}/a_{\uparrow\uparrow}}\right)$ \mbox{(dashed line) \cite{NoteSupplementary}.} The absence of collapse in our measurements shows the existence of a stabilizing beyond mean-field mechanism.

In the deeply bound regime the measured peak densities agree only qualitatively with the eGPE predictions, see left and right panels of Fig. 3(a). The discrepancies might stem from two sources. First, we have considered that the spin composition of the system adjusts to $N_{\uparrow}/N_{\downarrow}=\sqrt{a_{\downarrow\downarrow}/ a_{\uparrow\uparrow}}$ while we have seen experimentally that population imbalances are possible. Second, our decay model is very simplified and assumes that the $\ket{\downarrow}$ losses are immediately accompanied by the disappearance of $\ket{\uparrow}$ atoms when, in reality, these require a finite time to exit the observation region.

In a last series of experiments we explore the phase diagram by approaching the bi-stability region from the soliton regime, see left inset of Fig. 4(a). We prepare the system in the crossover region at $B\sim56.3$ G and hold it in the crossed optical dipole trap for a variable time (1 to 120 ms). Owing to three-body recombination, this results in atom numbers $N=3000$ to $7000$. We then remove the vertical trapping beam and rapidly decrease $B$ to its final value at a rate of $93.76$ G/s. At the boundaries of the bi-stable region, the density of the system becomes discontinuous. Experimentally, we observe that the self-bound state cannot adjust to this abrupt change and fragments, see right panel. A similar behavior is observed in harmonically trapped dipolar gases \cite{Kadau2016, Schmitt2017}. To locate the fragmentation point, we record the atom number in the initially self-bound region and observe an abrupt drop at a critical magnetic field. As shown in the left panel of Fig. 4(a), its value depends on the initial atom number. We summarize the position of the fragmentation point in the $N-B$ plane in Fig. 4(b).

To interpret our observations, we exploit again the variational model.  It predicts that, although in the bi-stable region both solitons and droplets exist, their energies coincide only along a transition line (solid line). Above (below) it, solitons (droplets) become metastable, and only disappear at the upper (lower) boundary (dashed lines). The three situations are depicted in the right panel of Fig. 4(b). Experimentally, we prepare the mixture in a regime where only solitons exist. Therefore, when entering the bi-stable region we expect it to follow preferentially the metastable soliton solution, with which it connects smoothly. At the upper boundary the metastable soliton disappears and only dense droplets are possible. Here fragmentation is expected to occur. Our experimental results support this hypothesis since, within error bars, they agree with the numerical eGPE predictions (colored area) without any fitting parameters.

In conclusion we have shown that an attractive mixture of BECs confined in an optical waveguide always hosts self-bound states, which correspond to composite bright solitons, quantum liquid droplets, or interpolate smoothly between both limits depending on the values of the atom number, interaction strength and confinement. We have characterized their spin composition and density, and mapped out the upper boundary of the bi-stable region separating solitons and droplets. Future experimental directions include the study of metastability and hysteresis when crossing the soliton-to-droplet transition from different directions, of collisions  between two self-bound states -- which are expected to display very different behavior in the soliton and droplet limits \cite{Nguyen2014,Adhikari2016,Ferrier2016b} -- and of finite temperature effects \cite{Boudjemaa2017} in spin imbalanced systems, where the excess component would provide a well controlled thermal bath.

\emph{Note: During completion of this work, we became aware of related experiments by the LENS group \cite{Fattori2017}.}

We thank A. Celi and P. Thomas for a careful reading of the manuscript, and P. Naylor for technical assistance with the numerical simulations. We acknowledge insightful discussions with G. Astrakharchik, J. Boronat, M. Fattori, I. Ferrier-Barbut, B. Malomed, D. Petrov, L. Santos, G. Semeghini and L. Torner. We thank A. Simoni and M. Tomza for calculations of the potassium scattering lengths. We acknowledge funding from Fundaci\'{o} Privada Cellex, EU (MagQUPT-631633 and QUIC-641122), Spanish MINECO (StrongQSIM FIS2014-59546-P and Severo Ochoa SEV-2015-0522), DFG (FOR2414), Generalitat de Catalunya (SGR874 and CERCA program), and Fundaci\'{o}n BBVA. PC acknowledges support from the Marie Sk{\l}odowska-Curie actions (TOPDOL-657439), CRC from CONACYT (402242/ 384738), JS from FPI (BES-2015-072186), and L. Tarruell from the Ram\'{o}n y Cajal program (RYC-2015-17890).

\section*{Supplementary material}

\subsection{A. Scattering lengths and inelastic losses}

\textit{Scattering lengths.} We perform the experiments with a mixture of $^{39}$K in states $\ket{\uparrow}\equiv\ket{F=1, m_F=-1}$ and $\ket{\downarrow}\equiv\ket{F=1, m_F=0}$. We explore the magnetic field range $B=55-57.5$ G, in the vicinity of a Feshbach resonance for state $\ket{\downarrow}$. This gives access to a situation where $a_{\uparrow\uparrow}, a_{\downarrow\downarrow}>0$ and $a_{\uparrow\downarrow}<0$. In this regime the repulsive intra-state and attractive inter-state scattering lengths almost completely cancel out, resulting in a small residual mean-field energy proportional to $\delta a=a_{\uparrow\downarrow}+\sqrt{ a_{\uparrow\uparrow} a_{\downarrow\downarrow}}$. Fig. S1 summarizes the values of $a_{\uparrow\downarrow}$, $a_{\uparrow\uparrow}$, $a_{\downarrow\downarrow}$ and $\delta a$ predicted by the $^{39}$K model interaction potentials of refs. \cite{D'Errico2007, Roy2013, Simoni2016}. We have verified that the model potentials of ref. \cite{Falke2008} yield equivalent results \cite{Tomza2016}.
\begin{suppfigure}[!h]
\centering
\includegraphics[clip]{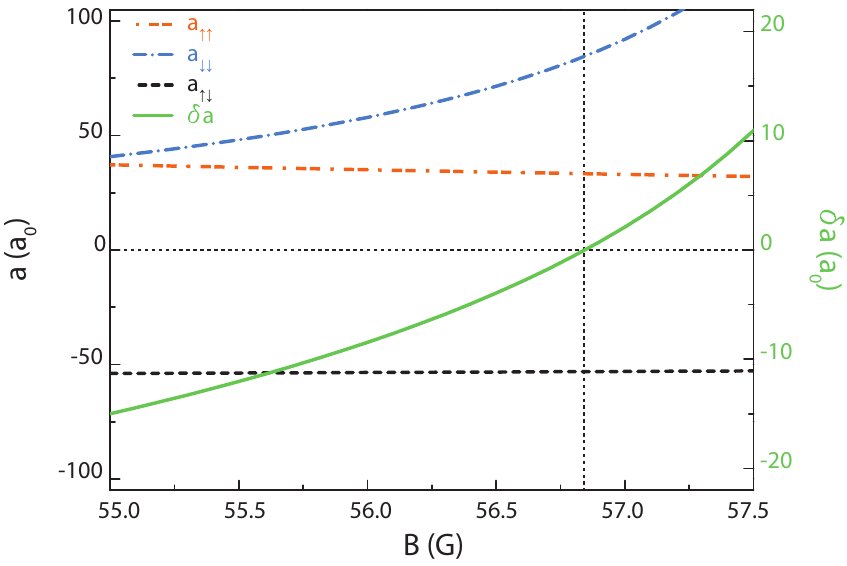}
\caption{Scattering lengths $a_{\uparrow\uparrow}$, $a_{\downarrow\downarrow}$, $a_{\uparrow\downarrow}$, and parameter $\delta a= a_{\uparrow\downarrow}+\sqrt{a_{\uparrow\uparrow}a_{\downarrow\downarrow}}$ (expressed in units of the Bohr radius $a_0$) as a function of magnetic field $B$ for a $^{39}$K  mixture in states $\ket{\uparrow}\equiv\ket{F=1, m_F=-1}$ and $\ket{\downarrow}\equiv\ket{F=1, m_F=0}$. }\label{figS1}
\end{suppfigure}

\emph{Inelastic losses. }In the same magnetic field range, the scattering model predicts two-body inelastic collision rates $K_{\uparrow\downarrow}<1.92\times10^{-16}$ cm$^3$/s, $K_{\uparrow\uparrow}<2.34\times10^{-15}$ cm$^3$/s, and $K_{\downarrow\downarrow}<7.28\times10^{-16}$ cm$^3$/s \cite{Simoni2016}, which stem from the dipole-dipole interaction.

Since no theoretical predictions are available for the three-body recombination rates $ K_{\uparrow\uparrow\uparrow}$, $K_{\uparrow\uparrow\downarrow}$,  $K_{\uparrow\downarrow\downarrow}$ and $K_{\downarrow\downarrow\downarrow}$, we determine their values experimentally. To this end, we trap thermal atomic clouds in a crossed optical dipole trap of mean trap frequency $\bar{\omega}/2\pi = 331(7)$ Hz and depth $U_0/k_{\mathrm{B}}=36(2)\,\mu$K, with $ k_{\mathrm{B}}$ the Boltzmann constant. We then record the time evolution of their atom number $N$ and temperature $T$. For single-component systems we model our measurements by the set of coupled equations
\begin{eqnarray}
\frac{\dot{N}}{N} &=& - \frac{\beta}{\sqrt{27}} \frac{K_3^{\mathrm{th}} N^2}{T^3}\label{N}\\
\frac{\dot{T}}{T} &=& \frac{\beta}{\sqrt{27}} \frac{K_3^{\mathrm{th}} N^2(T+T_h)}{3 T^4}\label{T},
\end{eqnarray}
which includes the effect of anti-evaporation. Here $\beta=(m\bar{\omega}^2/2\pi k_{\mathrm{B}})^{3/2}$, $K_3^{\mathrm{th}}$ denotes the corresponding three-body recombination rate and $T_h$ is a free parameter that takes into account recombination heating \cite{Weber2003}. We neglect evaporation effects, which is a good assumption in our parameter regime $T\lesssim 2.5\mu$K$\ll U_0/k_{\mathrm{B}}$.

We have performed these measurements in single-component samples of $\ket{\uparrow}$, $\ket{\downarrow}$ and in mixtures of different concentrations, see Fig. S2. For the magnetic field range of the experiment we find that losses of $\ket{\downarrow}$ dominate over all the other processes and that the effective three-body loss rate of the mixture is proportional to the fraction of atoms in this state. The inset of Fig. S2 summarizes the magnetic field dependence of  $K^{\mathrm{th}}_{\downarrow\downarrow\downarrow}$ as a function of magnetic field $B$. It remains approximately constant in the range $55.5-56.5$ G studied in Fig. 3 of the main text. We analyze the corresponding decay of the self-bound atom number using the average value $K^{\mathrm{th}}_{\downarrow\downarrow\downarrow} = 3(1)\times10^{-27}$ cm$^6$/s, see section C. The other three rates are compatible with the $^{39}$K background value $7.74\times10^{-29}$ cm$^6$/s \cite{Zaccanti2009, Lepoutre2016}. Note that all our three-body loss rate measurements have a large systematic uncertainty (not included in the error bar) of up to a factor of two, dominated by the $25\%$ systematic uncertainty of the atom number calibration.

\begin{suppfigure}[!h]
\centering
\includegraphics[clip]{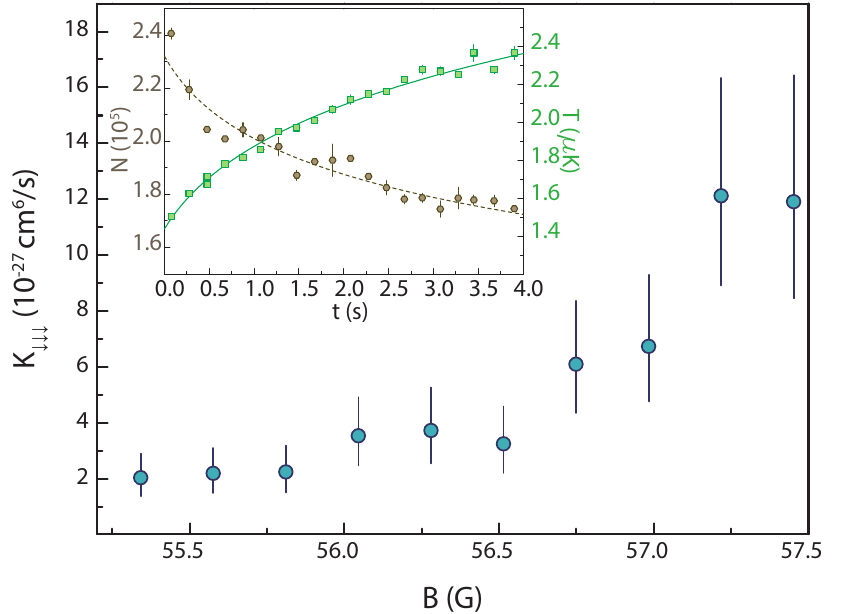}
\caption{Three-body recombination rate $K^{\mathrm{th}}_{\downarrow\downarrow\downarrow}$ as a function of magnetic field $B$. Inset: time evolution of the atom number $N$ and temperature $T$ at $B=55.811(9)$ G, fitted to eq. (\ref{N}) (dotted line) and eq. (\ref{T}) (solid line).}\label{figS2}
\end{suppfigure}

In conclusion for the densities explored in the experiment inelastic processes are dominated by three-body recombination in the $\downarrow\downarrow\downarrow$ channel.

\subsection{B. Theoretical analysis}

\emph{Theoretical model.} Following ref. \cite{Petrov2015} we describe the mixture with an effective low-energy theory: an extended Gross-Pitaevskii equation (eGPE) with an additional repulsive term. It includes the effect of quantum fluctuations as an effective potential for the low-energy degrees of freedom of the system. This approach is valid in the absence of spin excitations \cite{Petrov2015}. We fulfill this condition by assuming identical spatial modes for the two components $\big(\Psi_{\uparrow}=\sqrt{n_{\uparrow}}\phi$ and $\Psi_{\downarrow}=\sqrt{n_{\downarrow}}\phi\big)$ and the density ratio $n_{\uparrow}/n_{\downarrow}=\sqrt{a_{\downarrow\downarrow }/a_{\uparrow\uparrow}}$ \cite{Petrov2015, Petrov2016}. In our experimental conditions the individual BECs can be considered as three-dimensional, since the harmonic oscillator length $a_{\mathrm{ho}}$ exceeds their healing lengths by more than an order of magnitude. The corresponding energy density functional therefore reads
\begin{equation}\label{EeGPE}
\begin{split}
\mathcal{E}&=\mathcal{E}_{\mathrm{kin}}+\mathcal{E}_{\mathrm{trap}}+\mathcal{E}_{\mathrm{MF}}+\mathcal{E}_{\mathrm{LHY}}\\
&=\frac{\hbar^2}{2m}n_0\abs{\nabla\phi}^2+V_{\mathrm{trap}}n_0\abs{\phi}^2\\
&+\frac{4\pi\hbar^2\delta a}{m}\frac{\sqrt{a_{\downarrow\downarrow}/a_{\uparrow\uparrow}}}{(1+\sqrt{a_{\downarrow\downarrow}/a_{\uparrow\uparrow}})^2}n_0^2\abs{\phi}^4 \\
&+\frac{256\sqrt{\pi}\hbar^2}{15m}\bigg(\frac{n_0\sqrt{a_{\uparrow\uparrow}a_{\downarrow\downarrow}}}{1+\sqrt{a_{\downarrow\downarrow}/a_{\uparrow\uparrow}}}\bigg)^{5/2}f\bigg(\frac{a_{\uparrow\downarrow}^2}{a_{\uparrow\uparrow}a_{\downarrow\downarrow}},\sqrt{\frac{a_{\downarrow\downarrow}}{a_{\uparrow\uparrow}}}
\bigg)\abs{\phi}^5,\nonumber
\end{split}
\end{equation}
where $n_0 = n_{\uparrow}+n_{\downarrow}$ and $\mathcal{E}_{\mathrm{kin}}$,  $\mathcal{E}_{\mathrm{trap}}$, $\mathcal{E}_{\mathrm{MF}}$ and $\mathcal{E}_{\mathrm{LHY}}$ denote the kinetic, potential, mean-field and quantum fluctuation (Lee-Huang-Yang) contributions to the energy density of the mixture, respectively. Furthermore, $V_{\mathrm{trap}}=\frac{1}{2}m\omega^2 r^2$ corresponds to the harmonic confinement of the optical waveguide and $f(x,y)=\sum_{\pm}\left(1+y\pm\sqrt{(1-y)^2+4xy}\right)^{5/2}/4\sqrt{2}$ \cite{Larsen1963}.
This energy functional results in the extended Gross-Pitaevskii equation given in the main text
\begin{equation}\label{eGPEsupp}
i\hbar \dot{\phi}=\left[\left(-\frac{\hbar^2}{2m}\nabla^2+V_{\mathrm{trap}}\right)+\alpha n_0\abs{\phi}^2+\gamma n_0^{3/2}\abs{\phi}^{3}\right]\phi,
\end{equation}
where for clarity we have made explicit the density scaling of the two last terms by defining
\begin{eqnarray}
\alpha&=&\frac{8\pi\hbar^2\delta a}{m}\frac{\sqrt{a_{\downarrow\downarrow}/a_{\uparrow\uparrow}}}{(1+\sqrt{a_{\downarrow\downarrow}/a_{\uparrow\uparrow}})^2}\nonumber,\\
\gamma&=&\frac{128\sqrt{\pi}\hbar^2}{3m} \bigg(\frac{\sqrt{a_{\downarrow\downarrow}a_{\uparrow\uparrow}}}{1+\sqrt{a_{\downarrow\downarrow}/a_{\uparrow\uparrow}}}\bigg)^{5/2} f\bigg(\frac{a_{\uparrow\downarrow}^2}{a_{\uparrow\uparrow}a_{\downarrow\downarrow}},\sqrt{\frac{a_{\downarrow\downarrow}}{a_{\uparrow\uparrow}}}\bigg).\nonumber
\end{eqnarray}

\emph{Numerical analysis.} To obtain the phase diagram depicted in Fig. 3(a), we find numerically the stationary solutions of eq. (\ref{eGPEsupp}) using the three-dimensional MATLAB toolbox of ref. \cite{Antoine2014}. For each magnetic field $B$ we first solve the eGPE for $N=1000$, using as initial guess for $\phi$ a three-dimensional Gaussian of size $a_{\mathrm{ho}}$. We subsequently compute the solution for increasing values of $N$, choosing as initial guess the function $\phi$ determined in the previous step. Below $B_c\sim 55.85$ G we find two distinct solutions of very different peak densities $n_0$: a dilute bright soliton (S) at low $N$ and a dense liquid-like droplet (D) at high $N$. They coexist in a bi-stable region, where one of them is metastable. In this regime we perform two separate $N$ sweeps: increasing $N$ starting from the soliton regime yields the soliton solution, whereas decreasing $N$ starting from the droplet regime yields the droplet solution. To determine the boundaries of the bi-stable region we then compute $\left(n_0^D-n_0^S\right)/\left(n_0^D+n_0^S\right)$, see Fig. 4(b). For all parameters explored in the experiment, we find that $\phi$ is well approximated by a Gaussian.

\emph{Variational analysis.} We gain further insight on the properties of the system by performing a variational analysis, using the Gaussian ansatz $\phi=\mathrm{e}^{-r^2/2\sigma_r^2-z^2/2\sigma_z^2}$. It yields the following functional for the total energy of the mixture $E=\int \mathrm{d}\mathbf{r}\,\mathcal{E}$
\begin{eqnarray}\label{VarE}
\frac{E(\sigma_r,\sigma_z)}{N\hbar\omega}&=& \frac{1}{{N\hbar\omega}}\left(E_{\mathrm{kin}}+E_{\mathrm{trap}}+E_{\mathrm{MF}}+E_{\mathrm{LHY}}\right)\nonumber\\
   &=&\frac{1}{4}\left(\frac{2 a_{\mathrm{ho}}^2}{\sigma_r^2} + \frac{a_{\mathrm{ho}}^2}{\sigma_z^2}\right)+ \frac{1}{4} \left(\frac{2\sigma_r^2}{a_{\mathrm{ho}}^2}\right)\nonumber\\
   &+&\frac{1}{\sqrt{2\pi}}\frac{N a_{\mathrm{ho}}^3}{\sigma_r^2\sigma_z} \left(\frac{2 \sqrt{a_{\downarrow\downarrow}/a_{\uparrow\uparrow}}\delta a }{\left(1+\sqrt{a_{\downarrow\downarrow}/a_{\uparrow\uparrow}}\right)^2a_{\mathrm{ho}}}\right)\nonumber\\
   &+&\sqrt{\frac{2}{5}}\frac{512}{75\pi^{7/4}}\frac{N^{3/2}a_{\mathrm{ho}}^{9/2}}{\sigma_r^3\sigma_z^{3/2}}\left(\frac{\sqrt{a_{\uparrow\uparrow}a_{\downarrow\downarrow}}}{a_{\mathrm{ho}}(1+\sqrt{a_{\downarrow\downarrow}/a_{\uparrow\uparrow}})}\right)^{5/2}\nonumber\\
   &&f\left(\frac{a_{\uparrow\downarrow}^2}{a_{\uparrow\uparrow}a_{\downarrow\downarrow}},\sqrt{\frac{a_{\downarrow\downarrow}}{a_{\uparrow\uparrow}}}\right).
\end{eqnarray}

In the absence of quantum fluctuations this expression is equivalent to the one commonly employed to describe a single-component condensate of scattering length $a$ \cite{Pethick2008}, provided one does the replacement
\begin{equation}
 \frac{2 \sqrt{a_{\downarrow\downarrow}/a_{\uparrow\uparrow}}}{\left(1+\sqrt{a_{\downarrow\downarrow}/a_{\uparrow\uparrow}}\right)^2} \delta a\rightarrow a.\nonumber
\end{equation}
For $\delta a <0$ eq. \eqref{VarE} contains a composite bright soliton solution stabilized by the balance between $E_{\mathrm{kin}}>0$ and $E_{\mathrm{MF}}<0$. At the mean-field level, solitons only exist if the system is effectively one-dimensional. They therefore collapse when the mean-field attraction becomes comparable to the trapping energy $\hbar \omega$. The collapse criterion reads
\begin{equation}\label{SolitonCollapse}
N_c = 0.6268  \frac{\left(1+\sqrt{a_{\downarrow\downarrow}/a_{\uparrow\uparrow}}\right)^2a_{\mathrm{ho}}}{2 \sqrt{a_{\downarrow\downarrow}/a_{\uparrow\uparrow}}\abs{\delta a}},\nonumber
\end{equation}
where the pre-factor has been computed numerically from eq. (\ref{eGPEsupp}), since the variational calculation is known to overestimate the soliton stability \cite{Carr2002}.

Taking into account quantum fluctuations considerably modifies the behavior of the mixture. A second self-bound solution is then possible: a quantum liquid droplet stabilized by the balance between $E_{\mathrm{MF}}<0$ and $E_{\mathrm{LHY}}>0$. At low atom numbers, $E_{\mathrm{kin}}>0$ plays an important role as well, due to the scaling of the different energy contributions with $N$: $E_{\mathrm{kin}}\propto N$, $E_{\mathrm{MF}}\propto N^2$ and $E_{\mathrm{kin}}\propto N^{5/2}$. Below a critical atom number, it becomes sufficiently strong to unbind the droplets \cite{Petrov2015, Cabrera2017}.

In the deeply bound regime solitons and droplets are distinct solutions. They co-exist in a bi-stable region. There, the upper boundary corresponds to the collapse of solitons into droplets when the system no longer behaves as one-dimensional. The lower one indicates the dissociation of droplets into solitons due to kinetic effects. The distinction between solitons and droplets disappears when the free-space droplet size becomes comparable to the harmonic oscillator length $a_{\mathrm{ho}}$, which determines the radial soliton size. Both become then smoothly connected in a crossover. Thus, the position of the critical point separating the transition and crossover regimes is determined by the confinement of the optical waveguide.

\subsection{C. Data analysis}

\textit{Image analysis.} The \emph{in situ} images of the mixture are taken exploiting the polarization phase contrast scheme described in ref. \cite{Cabrera2017}. It allows to detect the two states with essentially the same sensitivity. Therefore, the images yield the total column density of the system $n_c(x,z)= n_{c,\uparrow}(x,z)+ n_{c,\downarrow} (x,z)$.

The vast majority of the measurements performed in the $\delta a<0$ regime correspond to self-bound states smaller or on the order of the imaging resolution. In Fig. 1 we fit the experimental density profiles with a two-dimensional Gaussian in order to extract their $1/\mathrm{e}$ width $\sigma_z$, which could be strongly affected by lens aberrations. Thus, it shall not be used to characterize the system quantitatively but only to indicate its self-bound character. Similarly, in Fig. 2 we extract the fraction of self-bound atoms from bi-modal Gaussian fits to the density profiles. However, these images are only exploited to determine the optimal rf pulse time and the spin composition of the mixture is obtained from time-of-flight images, which are not affected by the spatial resolution.

For the data presented in Figs. 3 and 4, we determine quantitatively the atom number of the self-bound states from the \emph{in situ} images by evaluating the zeroth moment of the images $N=M_{00}=\sum_{x,z} n_c(x,z)$, which is independent of lens aberrations \cite{Bradley1997}. In order to count only the self-bound atoms we crop the images around the maximal column density and extract $N$ from this observation region, see insets in Figs. 3 and 4. We have verified that increasing the crop size in the direction perpendicular to the waveguide does not modify the results. The longitudinal crop size needs to be adjusted more carefully to avoid counting excess atoms that are expanding in the waveguide. We fix its value by comparing the atom number extracted from \emph{in situ} images with no excess component with time-of-flight measurements. We find that for all the data, possible errors in $N$ associated to incorrect choices of the longitudinal crop size remain $<10\%$, below the systematic error of the $N$ calibration ($25\%$).

\textit{Decay model.} To circumvent the limitations imposed by the imaging resolution, in Fig. 3 we exploit the decay of the self-bound atom number caused by inelastic processes to determine the density of the system.  In the regime explored in the experiment, one- and two-body processes are negligible compared to three-body recombination. The latter is described by the three-body recombination rates \cite{PetrovSantos2017}
\begin{eqnarray}
K_{\sigma\sigma\sigma}&=&\frac{1}{3!}K^{\mathrm{th}}_{\sigma\sigma\sigma}
\left[1+\frac{6}{n_{\sigma}^2}\frac{\partial \mathcal{E}_{\mathrm{LHY}}}{\partial g_{\sigma\sigma}} \right],\nonumber\\
K_{\sigma\sigma\sigma'}&=&\frac{1}{2!}K^{\mathrm{th}}_{\sigma\sigma\sigma'}
\left[1+\frac{2}{n_{\sigma}^2}\frac{\partial \mathcal{E}_{\mathrm{LHY}}}{\partial g_{\sigma\sigma}}
+\frac{2}{n_{\sigma} n_{\sigma'}}\frac{\partial \mathcal{E}_{\mathrm{LHY}}}{\partial g_{\sigma\sigma'}}\right],\nonumber
\end{eqnarray}
where $\sigma,\sigma'$ denote the spin states, $g_{\sigma\sigma'}=4\pi \hbar^2 a_{\sigma\sigma'}/m$, and $K_3^{\mathrm{th}}$ are the thermal rates determined in section A. The numerical pre-factors result from the indistinguishability of bosonic atoms \cite{Kagan1985}, and the terms involving $\cal{E}_{\mathrm{LHY}}$ correspond to the beyond mean-field corrections to the three-body correlation functions of the mixture. In the regime explored in the experiment, they remain $<10\%$. Since this is well below the uncertainties of the thermal rates, we neglect them as in ref. \cite{Chomaz2016}.

Describing the decay of the self-bound states requires taking into account simultaneously two effects: (i) real loss of $\ket{\downarrow}$ atoms, since  $K_{\downarrow\downarrow\downarrow}$ is much larger than the three other rates; (ii) expulsion (and subsequent expansion along the waveguide) of $\ket{\uparrow}$ atoms, in order to maintain the optimal spin composition of the self-bound state. Modelling accurately the dynamics of these combined loss, expulsion and expansion processes goes beyond the scope of this work. Note that under these conditions it is not clear that the eGPE model (derived neglecting explicitly spin excitations \cite{Petrov2015}) or simple extensions of it remain valid. We instead simplify considerably the problem by assuming that $\ket{\downarrow}$ losses are instantaneously accompanied by the disappearance of $\ket{\uparrow}$ atoms required to maintain $N_{\uparrow}/N_{\downarrow}$ fixed in the self-bound state. The decay of the self-bound atom number is then given by the rate equation
\begin{equation}
\frac{\dot{N}}{N}=-K^{\mathrm{eff}}_3 \langle n^2\rangle,\label{K3eff}\nonumber
\end{equation}
where $\langle n^2\rangle=\frac{1}{N}\int \mathrm{d}\mathbf{r}\, n^3$ and the effective three-body loss coefficient is $K_3^{\mathrm{eff}}= {K_{\downarrow\downarrow\downarrow}}\Big/\left(1+\sqrt{\frac{a_{\downarrow\downarrow}}{a_{\uparrow\uparrow}}}\right)^{2}.$

To extract $\dot{N}/N$ from the decay curves, we fit them with the empirical function $$N(t)=N_{\infty}+p N_0\mathrm{e}^{-(t-T_0)/T_1}+(1-p)N_0\mathrm{e}^{-(t-T_0)/T_2},$$ where $N_0$, $N_{\infty}$, $T_0$, $T_1$, $T_2$ and $p$ are free parameters.
We finally determine the peak density of the system from
\begin{equation}
n_0=3^{3/4}\sqrt{\langle n^2\rangle}=3^{3/4}\left(1+\sqrt{\frac{a_{\downarrow\downarrow}}{a_{\uparrow\uparrow}}}\right)\sqrt{\frac{1}{K_{\downarrow\downarrow\downarrow}}\abs{\frac{\dot{N}}{N}}}.\nonumber
\end{equation}
Here, we have assumed a Gaussian density profile to relate the peak and average densities to facilitate the comparison to the theoretical model.

We have verified that the results obtained using a different experimental fitting function are well below the uncertainties introduced by the $K_{\downarrow\downarrow\downarrow}$ systematic error. In any case, we expect our determination of the density to be dominated by the simplifications of the decay model. Considering only the effect of $\ket{\downarrow}$ losses would reduce the determined densities by a factor of $2$.

\textit{Error analysis.} In Fig. 2(b), the optimal time $\tau_{\mathrm{op}}$ is extracted from a lorentzian fit to the $N_{\mathrm{SB}}/N$ curve. Its error $\Delta \tau$ corresponds to the standard error of the mean extracted from the fit. To extract the optimal spin ratio from Fig. 2(b), after determining $N_{\uparrow}/N_{\downarrow}$ from time-of-flight images, we fit it in the vicinity of $\tau_{\mathrm{op}}$ with a second order polynomial that we evaluate at $\tau_{\mathrm{op}}$. To extract the error of the ratio we evaluate the prediction bounds of the fit at $\tau_{\mathrm{op}}\pm\Delta \tau$ considering a confidence interval of $\sigma$.

In Fig. 4(b) the critical magnetic field for fragmentation $B_c$ corresponding to the initial atom number $N_i$ is obtained by fitting curves analogous to those of Fig. 4(a) with an error function $$N_{\mathrm{crop}} = \left(\frac{N_i-N_f}{2}\right) \mathrm{erf}\left(-\frac{B-B_c}{\sqrt{2} \sigma}\right)+ N_f,$$
where $N_i$, $N_f$, and $B_c$ are free parameters. The horizontal error bars of the fragmentation points correspond to $\sigma$, and the vertical ones to the $25\%$ systematic error on the atom number calibration.


\begin{thebibliography}{99}
\bibitem{DropletDimensionalityNote} For our experimental parameters, $a_{\mathrm{ho}}$ exceeds the healing length of each BEC by more than an order of magnitude.
\end{thebibliography}
\end{document}